\documentclass[aps,prd,twocolumn,superscriptaddress,showkeys,showpacs]{revtex4}
\usepackage{pdfsync}
\usepackage{mathptmx}
\usepackage{epsfig,epsf}
\usepackage{amsmath}
\usepackage{amsthm}
\usepackage{amsfonts}
\usepackage{amssymb}
\usepackage{dsfont}

\usepackage{slashed}

\newcommand \vev [1] {\langle{#1}\rangle}

\newcommand \widebar [1] {\overline{#1}}

\begin{document}


\title{
\hspace{15cm}
\phantom{ Baryon octet distribution amplitudes in Wandzura-Wilczek }
{ \textmd{DESY 15-254}}
\\[3mm]
Baryon octet distribution amplitudes in Wandzura-Wilczek approximation}

\date{\today}

\author{ I.~V.~Anikin}
\affiliation{Bogoliubov Laboratory of Theoretical Physics, JINR, 141980 Dubna, Russia}
\affiliation{Institut f\"ur Theoretische Physik, Universit\"at
   Regensburg, D-93040 Regensburg, Germany}
\author{ A.~N.~Manashov}
\affiliation{Institut f\"ur Theoretische Physik, Universit\"at
   Hamburg, D-22761 Hamburg, Germany}
\affiliation{Institut f\"ur Theoretische Physik, Universit\"at
   Regensburg, D-93040 Regensburg, Germany}

\date{\today}

\begin{abstract}
We study higher twist distribution amplitudes for the $SU_F(3)$ baryon octet. We identify independent functions
for all baryons in the isospin symmetry limit and calculate the Wandzura-Wilczek contributions to the twist-$4$
and $5$  distributions amplitudes.
\end{abstract}


\maketitle

%
\section{Introduction}

Hard exclusive processes give us  unique  possibility to study  internal structure of hadrons. The theoretical
description of exclusive processes is based on
 the QCD factorization
approach~\cite{Chernyak:1977as,Chernyak:1977fk,Chernyak:1987nu,Efremov:1979qk,Efremov:1978rn,Lepage:1979zb}.
Scattering amplitudes (decay amplitudes) in this approach are given by  convolution of the coefficient function
which can be calculated perturbatively with  nonpertubative functions -- the distribution amplitudes (DAs).
In the infinite momentum frame DAs can be interpreted as the  momentum fraction distributions of partons in hadrons.

The DAs are usually classified according to their twist. In the QCD factorization approach, where the relevant
$Q^2$ is large, the dominant contributions to an amplitude come from  DAs
 of lowest possible twist.
The higher twist DAs give rise to the power suppressed corrections. The factorization approach works quite well for
the mesons but for baryons it encounters conceptual difficulties (see
Refs.~\cite{Duncan:1979hi,Duncan:1979ny,Milshtein:1981cy,Milshtein:1982js,Kivel:2010ns}). Also one faces the
difficulties in attempt to provide a quantitative description of the current experimental data,  the electromagnetic
nucleon form factors, in particular.

A quantitative description of the nucleon electromagnetic form factors has been achieved in  the framework of the
light-cone rules  (LCSR)~\cite{Balitsky:1986st,Balitsky:1989ry,Chernyak:1990ag} by taking into account the power
suppressed \cite{Braun:2001tj,Braun:2006hz,Lenz:2003tq,Aliev:2008cs} and next-to-leading order
\cite{PassekKumericki:2008sj,Anikin:2013aka} corrections. As it has been shown, the power suppressed corrections,
which are parameterized by the higher twist nucleon DAs, give sizeable contribution for  moderate $Q^2\sim 2-5\,
\text{GeV}$.

Unfortunately, our knowledge of the nucleon (baryon) DAs is quite limited. Only the leading twist  nucleon DA is
known with some degree of certainty, while the estimates of the higher twist nucleon DAs are very poor  (see e.g.
Refs.~\cite{Braun:2000kw,Braun:2006hz,Lenz:2003tq,Anikin:2013aka} and reference therein).

At the same time, the higher twist DAs contain the contributions that are related to the lower twist DAs -- the
so-called Wandzura-Wilczek (WW) contributions~\cite{Wandzura:1977qf}. For mesons, the {\it genuine} higher twist DAs
often appear to be much smaller than the corresponding WW terms. In many  cases, keeping  only the WW terms  one
gets a quite good approximation for the higher twist DAs (the so-called WW approximation).

The lowest twist  meson DAs are defined by matrix elements of two-particle (quark-antiquark or gluon) light-ray
operators. For such operators the WW contributions were calculated a long ago. A detailed discussion of the method
can be found in Ref.~\cite{Ball:1998ff}. We also mention here  that
 WW corrections to the generalized parton distributions were derived
in~\cite{Belitsky:2000vx,Radyushkin:2000jy,Kivel:2000rb,Anikin:2001ge,Geyer:2004bx}.

The situation with baryon DAs is more complicated because they are determined by matrix elements of three-particle
(three-quarks) operators. Until now the WW corrections were only known for the first few moments of the nucleon
DAs~\cite{Braun:2001tj,Braun:2006hz}.

The effective technique that allows one to calculate the WW terms  for the multiparticle DAs was developed
in~\cite{Braun:2008ia,Anikin:2013yoa}. The approach is based on the spinor formalism and conformal wave expansion
for the light-ray operators. Using this technique  the WW corrections to the three-particle nucleon DAs were calculated
up to twist-$5$~\cite{Braun:2008ia,Anikin:2013yoa}. In the present paper we, keeping in mind the recent progress in
lattice calculations of the baryon DA moments~\cite{Bali:2015ykx}, derive   the $SU_F(3)$ baryon octet DAs in the WW
approximation.

The paper is organized as follows: in Sect.~\ref{sect:convention} we remind the basics of the spinor formalism and
fix our notations. In Sect.~\ref{sect:DAs}, we give the definitions of the DAs for the baryon octet. The
subsections~\ref{sect:mixed-tw3}, \ref{sect:mixed-tw45} contain the analysis of the mixed chirality DAs up to
twist-$5$ and in the subsection~\ref{sect:chiral-tw45} we consider the chiral DAs of twist-$4$ and $5$. The
Appendices contain the
$SU_F(3)$ relations between different DAs and  explicit expressions for the few first  polynomials entering the expansion
of baryon DAs.

\section{Spinor conventions}\label{sect:convention}
We closely follow the notations of Ref.~\cite{Braun:2011aw}. In the Weyl representation, the Dirac $\gamma$-matrices
take the following form
\begin{align}
\gamma^\mu=
 \begin{pmatrix}
 0 & \sigma^\mu\\
 \widebar\sigma^\mu & 0
 \end{pmatrix},
\end{align}
where
\begin{align}
\sigma^\mu=(I, \vec{\bf\sigma}), && \overline\sigma^\mu=(I, -\vec{\bf\sigma})\,.
\end{align}
and $\vec{\sigma}$ are the  Pauli matrices. The $\gamma_5$ matrix and the charge conjugation matrix
$C=i\gamma^2\gamma^0$ are defined as
\begin{align}
 \gamma_5= \begin{pmatrix}
                 - I  & 0\\
                   0 & I
             \end{pmatrix},
             &&
C=\begin{pmatrix}
                 i\sigma^2 &0\\
                 0&-i\sigma^2
   \end{pmatrix}\,.
\end{align}
The Dirac spinor is constructed from two Weyl spinors:
\begin{eqnarray}
\label{spinor2}
q=
 \begin{pmatrix}
  q^\downarrow\\
  q^\uparrow
 \end{pmatrix}
 =
 \begin{pmatrix}
  \psi_{\alpha}\\
  \bar\chi^{\dot{\alpha}}
 \end{pmatrix}.
\end{eqnarray}
Here $q^{\downarrow(\uparrow)}=\frac12(1\mp \gamma_5) q$  are the left (right) handed quark fields,  respectively.

An arbitrary vector $a_\mu$ can be mapped to a $2\times 2$ matrix as
\begin{align}
a_{\alpha\dot\alpha}=a_\mu \sigma^\mu_{\alpha\dot\alpha}=\begin{pmatrix}
a_0-a_3 & -a_1+i a_2\\
-a_1-ia_2 & a_0+a_3
\end{pmatrix}.
\end{align}
In the  studies of hard processes, the light-like vectors, $n$ and $\bar n$ ($n^2=\bar n^2=0,$ $(n\cdot\bar n)\neq
0$), are usually introduced. They can be parameterized  by two auxiliary Weyl spinors $\lambda$ and $\mu$ as
\begin{align}
\label{lcb-matrix}
n_{\alpha\dot{\alpha}} = \lambda_\alpha\bar\lambda_{\dot{\alpha}}, &&
\tilde n_{\alpha\dot{\alpha}}= \mu_\alpha\bar\mu_{\dot{\alpha}}.
\end{align}
The rules for rising (lowering) spinor indices read
\begin{align*}
\lambda^\alpha=\epsilon^{\alpha\beta}\lambda_\beta, &&
\lambda_\alpha=\lambda^\beta \epsilon_{\beta\alpha},
 &&\bar\lambda^{\dot\alpha}=\bar\lambda_{\dot\beta}\bar\epsilon^{\dot\beta\dot\alpha},&&
\bar\lambda_{\dot\alpha}=\bar\epsilon_{\dot\alpha\dot\beta}\bar\lambda^{\dot\beta},
\end{align*}
where the antisymmetric Levi-Civita tensors are normalized as
$\epsilon_{12}=\epsilon^{12}=-\bar\epsilon_{\dot1\dot2}=-\bar\epsilon^{\dot1\dot2}=1$.
The products of  Weyl spinors are defined as
\begin{align}
\label{scalar-prod}
(\lambda \mu)=\lambda^\alpha \mu_\alpha=-(\mu\lambda), &&
(\bar \lambda \bar \mu)=\bar \lambda_{\dot{\alpha}} \bar \mu^{\dot{\alpha}}=
-(\bar\mu\bar\lambda).
\end{align}
In the following we use shorthand notation for the quark fields projections  onto the auxiliary spinors:
\begin{flalign}\label{prq}
q^\downarrow_+=(\lambda q^\downarrow), \hskip 2mm q^\downarrow_-=(\mu q^\downarrow), \hskip 2mm
q^\uparrow_+=(\bar \lambda q^\uparrow), \hskip 2mm q^\uparrow_-=(\bar \mu q^\uparrow) &.
\end{flalign}
The quark fields can be written in the form
\begin{align}
\label{Spinor-Expand}
(\mu\lambda)\, q^\downarrow_\alpha = \lambda_\alpha q^\downarrow_- - \mu_\alpha q^\downarrow_+,
&&(\bar\lambda \bar\mu) q^\uparrow_{\dot{\alpha}} =
\bar\lambda_{\dot{\alpha}} q^\uparrow_- - \bar\mu_{\dot{\alpha}} q^\uparrow_+\,.
\end{align}
Constructing the light-ray operators it is useful to ascribe two quantum numbers -- twist and helicity -- to the
quark field projections~\eqref{prq}. The plus components, $q^{\downarrow(\uparrow)}_+$, have the {\it collinear}
twist equal to $1$, while the minus components,  $q^{\downarrow(\uparrow)}_-$, have twist $2$. The helicity of the
$q^\downarrow_+,q^\uparrow_-$ projections is $-\frac12$, whereas $q^\uparrow_+,q^\downarrow_-$ have  helicity $+\frac12$,
see Ref.~\cite{Braun:2008ia} for details.

\section{Baryon octet distribution amplitudes}\label{sect:DAs}

\subsection{Twist three distribution amplitudes}\label{sect:mixed-tw3}
In this paper we study  the three-quark distribution amplitudes of the baryon octet,
\begin{eqnarray}
\label{Octet-matrix}
B
=
\begin{pmatrix}
 \frac{\Lambda^0}{\sqrt{6}}+ \frac{\Sigma^0}{\sqrt{2}} &  \Sigma^+ &  p \\[2mm]
  \Sigma^- & \frac{\Lambda^0}{\sqrt{6}}- \frac{\Sigma^0}{\sqrt{2}} & n \\[2mm]
 \Xi^- &  \Xi^0  & -\frac{2}{\sqrt{6}} \Lambda^0
 \end{pmatrix}\,,
\end{eqnarray}
in the isospin symmetry limit. It is helpful to recall  the quark content of the isospin
multiplets:
$(p,n)\sim (uud, udd)$, $(\Sigma^+,\Sigma^0,\Sigma^-)\sim (uus,uds, dds)$,
$(\Xi^0,\Xi^-)\sim (uss,dss)$ and  $\Lambda^0\sim uds$.

The leading twist distribution amplitudes are defined  by the matrix elements of the twist-$3$ three-quarks
operator:
\begin{align}\label{tw-3-1}
\mathds{O}^\downarrow_{3,
\vec{m}}(\mathbf{z})= \epsilon^{ijk} q^{\downarrow,i}_{+,m_1}(z_1)\,
 q_{+,m_2}^{\uparrow,j}(z_2)\,  q^{\downarrow,k}_{+,m_3}(z_3).
\end{align}
Here $\mathbf{z}=\{z_1,z_2,z_3\}$, $\vec{m}=\{m_1,m_2,m_3\}$. $m_k$ are the flavor indices,
$m_k=u,d,s$ and $i,j,k$ are the color indices.
 The matrix element of~\eqref{tw-3-1} takes the form
\begin{align}\label{defPhi3}
\vev{0|
\mathds{ O}^\downarrow
_{3,\vec{m}}(\mathbf{z})|B}
=-\frac{p_+}2 (u_B)^{\downarrow}_{+}\int \mathcal{D}x  e^{-i p_+\sum\,
{x_i z_i}}\,\Phi_{3,\vec{m}}^{(B)}(x)\,,
\end{align}
where $\Phi_{3,\vec{m}}^{(B)}(x)$ is the corresponding DA,  $u_B$ stands for the  Dirac spinor of the  baryon $B$,
$p$ denotes its momentum, \mbox{$p_+=(p n)$}
 and the
integration measure is defined as follows
\begin{eqnarray}
\mathcal{D}x=dx_1\,dx_2\,dx_3\,\delta(1-x_1-x_2-x_3).
\end{eqnarray}
It is clear that the matrix element is nonzero only if the flavor indices of the operator match the flavor content
of the baryon. Invariance of the operator under the permutation $(z_1,m_1)\leftrightarrow (z_3,m_3)$  together with
the isospin symmetry allows one to express all DAs in terms of seven independent functions. We choose them as
\begin{equation}\label{Our-Fun}
\begin{tabular}{c c c c}
\hline\\[-3mm]
\hspace{10pt} $    \Phi_{3,\,uud}^{(p)}$ \hspace{10pt}&
                  \hspace{10pt} $    \Phi_{3,\,uus}^{(\Sigma^+)}$ \hspace{10pt}&
                          \hspace{10pt} $   \Phi_{3,\,ssu}^{(\Xi^0)}$ \hspace{10pt}&
                          \hspace{10pt} $    \Phi_{3,\,uds}^{(\Lambda)}$ \hspace{10pt}
\\[2mm]
\hline\\[-3mm]

           \hspace{10pt}  \hspace{10pt}&
                  \hspace{10pt} $    \Phi_{3,\,usu}^{(\Sigma^+)}$ \hspace{10pt}&
                          \hspace{10pt} $   \Phi_{3,\,sus}^{(\Xi^0)}$ \hspace{10pt}&
                          \hspace{10pt} $    \Phi_{3,\,usd}^{(\Lambda)}$ \hspace{10pt}
\\[2mm]
\hline
\end{tabular}
\end{equation}
We mark the DAs by their positions in this table and do not specify the flavor indices explicitly,
i.e. $\Phi_3^{(B,k)}(x)$ denote the DAs of the baryon $B$ in the $k$-th row ($k=1,2$).

One can easily see that the functions in the second row have  certain parity with respect to $x_1\leftrightarrow
x_3$ permutations. Namely, the DAs
$\Phi_{3}^{(\Sigma^+,2)}(x)$ and $\Phi_{3}^{(\Xi^0,2)}(x)$ are symmetric while $\Phi_{3}^{(\Lambda,2)}(x)$
is antisymmetric under this permutation. The DAs in the first row,  $\Phi_3^{(B,1)}(x)$,  do not possess any
symmetries under permutations of the arguments. The proton DAs, $\Phi_3^{(p,2)}=\Phi_{3,udu}^{(p)}$, in the second
row is absent because this function is not independent and can be expressed in term of $\Phi_3^{(p,1)}$, namely
\begin{eqnarray}
\label{Phi-N}
\Phi_3^{(p,2)}(x)=-\Phi_3^{(p,1)}(x_1,x_2,x_3)-\Phi_3^{(p,1)}(x_3,x_2,x_1).
\end{eqnarray}

The DAs~\eqref{Our-Fun} are related to the vector, axial-vector and tensor twist three DAs, $V_1,A_1,T_1$ which
appear in the decomposition of the three-quark operator with open spinor indices~\cite{Braun:2000kw}
\begin{multline}\label{VAT}
4\vev{0|\epsilon^{ijk} q_{m_1\alpha}^i(z_1n) q_{m_2\beta}^j(z_2n) q_{m_3\gamma}^k(z_3n)|B(p)}= \\
=v_{\alpha\beta, \gamma}\,V_1^B(\vec{z}) + a_{\alpha\beta, \gamma}\,A^B_1(\vec{z}) +
t_{\alpha\beta, \gamma}\,T_1^B(\vec{z})+\ldots,
\end{multline}
where the dots stand for higher twist terms and
\begin{equation}
\label{Notations}
\begin{split}
v_{\alpha\beta, \gamma} & = (\slashed{p}C)_{\alpha\beta} \,(\gamma_5 u^+(P))_\gamma,
\\
a_{\alpha\beta, \gamma} & = (\slashed{p}\gamma_5 C)_{\alpha\beta} \, (u^+(P))_\gamma,
\\
t_{\alpha\beta, \gamma} & = (i\sigma^{\perp}_{\mu p}C)_{\alpha\beta}\, (\gamma_\perp^\mu \gamma_5\, u^+(P))_\gamma.
\end{split}
\end{equation}
As usual, the DAs  in momentum space are defined by the Fourier transform
\begin{align}
F(\vec{z})=\int \mathcal{D}x \, e^{-i(pn)\,\sum
{x_i z_i}}\,F (x)\,.
\end{align}
Projecting both sides of  Eq.~(\ref{VAT}) onto the auxiliary spinors $\lambda$, $\bar\lambda$ one derives
\begin{eqnarray}
\label{Phi3-B1}
\Phi_3^{(B,1)}(x)=V^{B}_1(x)-A^{B}_1(x),
\end{eqnarray}
and (cf. Eq.~(\ref{Phi-N}))
\begin{eqnarray}
\label{T-B2}
\Phi_{3}^{(B,2)}(x_1,x_2,x_3)=-2T_1^{B}(x_1,x_3,x_2)
\,.
\end{eqnarray}
The conformal wave expansion of the DAs~\eqref{Our-Fun}  goes  over polynomials $P_{Nq}$ which are solutions of the
renormalization group equations,
\begin{align}\label{Phi3pm}
\Phi_{3}^{(B,k)}(x,\mu)=x_1x_2x_3\sum_{N,q} 
\phi^{(B, k)}_{Nq}(\mu) \,P_{Nq}(x)\,,
\end{align}
where $\mu$ is the renormalization scale. The common prefactor  is fixed by the conformal spins of the fields. The
functions
$P_{Nq}(x)$ are the homogeneous polynomials of degree $N$,
$P_{Nq}(s x)=s^N P_{Nq}(x)$,
that form an orthogonal system
\begin{align}\label{}
\int \mathcal{D}x\, x_1 x_2 x_3\,P_{Nq}(x)P^\dagger_{Nq'}(x)=\delta_{qq'}c_{Nq}^{-1}\,.
\end{align}
The index $q$ enumerates different polynomials of the same degree. The polynomials $P_{Nq}$ can be obtained as
solutions of one-loop RG equation for twist-$3$ three quark operators. Note that these polynomials do not depend on
the flavor content of three-quark operator because the  evolution kernels are flavor-blind. Each polynomial $P_{Nq}$
is either symmetric or antisymmetric under permutations of the first and third arguments,
\begin{align}\label{PNq}
P_{Nq}^{(\pm)}(x_1,x_2,x_3)= \pm P^{(\pm)}_{Nq}(x_3,x_2,x_1)\,.
\end{align}
 The reduced matrix element
$\phi^{(B,k)}_{Nq}$ can be expressed as a convolution integral
\begin{align}\label{}
\phi^{(B,k)}_{Nq}=c_{Nq}\int \mathcal{D}x\,P^\dagger_{Nq}(x)\,\Phi_{3}^{(B,k)}(x)\,.
\end{align}
To the one-loop accuracy, the reduced matrix elements $\phi_{Nq}(\mu)$ have an autonomous scale dependence.
Evidently, for the DAs $\Phi_{3}^{(\Xi^0,2)}$, $\Phi_{3}^{(\Sigma^+,2)}$ and $\Phi_{3}^{(\Lambda,2)}$ the sum
in~\eqref{Phi3} goes over the symmetric (antisymmetric) polynomials only. The first few polynomials with the
corresponding anomalous dimensions are given in Appendix~\ref{Appendix:Polynomials}.
More details can be found in  Refs.~\cite{Bergmann:1999ud,Stefanis:1999wy,Braun:1999te,Braun:2008ia}.

For later convenience we split the DAs $\Phi_3^{(B,1)}(x)$ into symmetric and antisymmetric parts with respect to
$x_1\leftrightarrow x_3$ permutation,
\begin{align}
\Phi_3^{(B,1)}(x)=\Phi_3^{(B,1+)}(x) + \Phi_3^{(B,1-)}(x)\,
\end{align}
and introduce the notation $\Phi_3^{(B,A)}(x)$, where the label $A$ takes  the values $A=\{1\pm,2\} $.
Let us assign the signature factor $\varkappa(B,A)$:
\begin{align}\label{signaturefactor}
\varkappa(B,1\pm)&=\pm\,, &&\varkappa(B\neq \Lambda,2)
 =+\,, && \varkappa(\Lambda,2)=-\,
\end{align}
to each DA.
The signature factor defines  parity of the polynomials $P_{Nq}^{\pm}$ that enter the conformal expansion for the corresponding DA, namely,
\begin{align}\label{Phi3}
\Phi_{3}^{(B,A)}(x,\mu)=x_1x_2x_3\sum_{N,q} 
\phi^{(B, A)}_{Nq}(\mu) \,P_{Nq}^{\varkappa(B,A)}(x)\,.
\end{align}
As we see in the next section the higher twist DAs can be written in quite similar form.

\subsection{Mixed chirality higher twist distribution amplitudes}\label{sect:mixed-tw45}
We define the higher twist three-quark operators of mixed chirality as
\begin{align}
\label{tw-4-1}
\mathds{O}^\uparrow_{4,\vec{m}}(\mathbf{z}) &=
  q^\downarrow_{+,m_1}(z_1)  q^\uparrow_{+,m_2}(z_2)  q^\downarrow_{-,m_3}(z_3),
\notag\\
\mathds{O}^\downarrow_{5,\vec{m}}(\mathbf{z}) &=
  q^\downarrow_{-,m_1}(z_1)  q^\uparrow_{-,m_2}(z_2)  q^\downarrow_{+,m_3}(z_3),
\end{align}
where it is tacitly assumed  that color indices are contracted with the antisymmetric tensor. We also recall  that
$q^\downarrow_-\equiv (\mu q^\downarrow )$, $q^\uparrow_-\equiv (\bar \mu q^\uparrow )$.
The operators
$\mathds{O}^{\uparrow}_{4,\vec{m}}$ and $\mathds{O}^\downarrow_{5,\vec{m}} $ have  helicities
$+1/2$ and $-1/2$, respectively.

 In the proton case  matrix elements of these operators define  the functions $\Phi_4,\Psi_4$  and
$\Phi_5,\Psi_5$, \cite{Braun:2000kw}.  For all other baryons in the octet (except neutron) there are three independent DAs related
to the matrix elements  of these operators.
Twist-$4$ and $5$ DAs can be defined as
\begin{align}\label{defPhi4}
\vev{0|\mathds{O}^\uparrow_{4,\vec{m}}(\mathbf{z})|B}
 & =\frac{m_B}4(\mu\lambda)  (u_B^{\uparrow})_+
\int \mathcal{D}x \,e^{-ip_+\sum z_k x_k}\,\Phi_{4,\vec{m}}^{(B)}(x)\,,
\notag\\
\vev{0|\mathds{O}^\downarrow_{5,\vec{m}}(\mathbf{z})|B}
 & =-\frac{m_B}4(\mu\lambda)  (u_B^{\uparrow})_-
\int \mathcal{D}x \,e^{-ip_+\sum z_k x_k}\,\Phi_{5,\vec{m}}^{(B)}(x)\,.
\end{align}
There are altogether  $11$  independent DAs for the $p$, $\Sigma^+, \Xi^0$ and $\Lambda$ baryons for each twist
$t=4,5$:
\begin{equation}\label{Fun-45}
\begin{tabular}{c c c c}
\hline\\[-3mm]
\hspace{10pt} $    \Phi_{t,\,uud}^{(p)}$ \hspace{10pt}&
                  \hspace{10pt} $    \Phi_{t,\,uus}^{(\Sigma^+)}$ \hspace{10pt}&
                          \hspace{10pt} $   \Phi_{t,\,ssu}^{(\Xi^0)}$ \hspace{10pt}&
                          \hspace{10pt} $    \Phi_{t,\,uds}^{(\Lambda)}$ \hspace{10pt}
\\[2mm]

           \hspace{10pt}  \hspace{10pt}&
                  \hspace{10pt} $    \Phi_{t,\,usu}^{(\Sigma^+)}$ \hspace{10pt}&
                          \hspace{10pt} $   \Phi_{t,\,sus}^{(\Xi^0)}$ \hspace{10pt}&
                          \hspace{10pt} $    \Phi_{t,\,usd}^{(\Lambda)}$ \hspace{10pt}

\\[2mm]
 \hspace{10pt} $    \Phi_{t,\,duu}^{(p)}$ \hspace{10pt}&
                  \hspace{10pt} $    \Phi_{t,\,suu}^{(\Sigma^+)}$ \hspace{10pt}&
                          \hspace{10pt} $   \Phi_{t,\,uss}^{(\Xi^0)}$ \hspace{10pt}&
                          \hspace{10pt} $    \Phi_{t,\,sdu}^{(\Lambda)}$ \hspace{10pt}
\\[2mm]
\hline
\end{tabular}
\end{equation}
Similar to the twist-$3$ case we introduce the notation $\Phi_t^{(B,A)}(x)$,  with the index $A$ taking the values $(1\pm,2)$.
The function  $ \Phi_t^{(B,2)}(x)$ stands for the DA in the second row of the table~\eqref{Fun-45} while
$\Phi_t^{(B,1\pm)}$ is given by the combination of DAs in the first and last row, namely
\begin{align}
\Phi_t^{(B,1\pm)}(x)=\frac12 \Big(\Phi_t^{(B,1)}(x)\pm \Phi_t^{(B,3)}(x)\Big)\,.
\end{align}
We want to stress here that
contrary to the twist-$3$ functions, the higher twist functions, $\Phi_t^{(B,A)}(x)$, are not symmetric under the arguments permutation.

In this notation the conventional proton DAs of twist $4$ and $5$~\cite{Braun:2000kw} take the following form
\begin{align}
\Phi_{4(5)}(x) & = \Phi_{4(5)}^{(p,1+)}(x_1,x_2,x_3)+\Phi_{4(5)}^{(p,1-)}(x_1,x_2,x_3)\,,\notag\\
\Psi_{4(5)}(x) & = \Phi_{4(5)}^{(p,1+)}(x_3,x_1,x_2)-\Phi_{4(5)}^{(p,1-)}(x_3,x_1,x_2)\,.
\end{align}

The higher twist DAs $\Phi_{t}^{(B,A)}(x)$ contain  contributions from both the operators of the {\it geometrical}
twist-$t$ and the  lower twist operators. Therefore, we represent the higher twist DAs in the following form
\begin{flalign}\label{Phi4}
\Phi_{4}^{(B,A)}(x)& = \Phi_{4}^{(B,A),g}(x)+\Phi_{4}^{(B,A),WW_3}(x)\,,&\\
\Phi_{5}^{(B,A)}(x)& = \Phi_{5}^{(B,A),g}(x)+\Phi_{5}^{(B,A),WW_3}(x)+\Phi_{5}^{(B,A),WW_4}(x).\notag&
\end{flalign}
The conformal wave expansion of the genuine twist-$4$ DAs, $\Phi_{4}^{(B,A),g}(x)$, takes the form
\begin{align}\label{Phi5}
\Phi_{4}^{(B,A),g}(x) & = x_1x_2\sum_{Nq} \eta_{Nq}^{(B,A)}(\mu) \, R^{\varkappa(B,A)}_{Nq}(x)\,,
\end{align}
where   the signature factor $\varkappa(B,A)$  is defined in~\eqref{signaturefactor}.

 The polynomials  $R_{Nq}^{(\pm)}$ are the
eigenfunctions of the evolution kernel for twist-$4$ operators, see Ref.~\cite{Braun:2008ia} for a general
discussion. Index $N$ stands for a degree of the polynomial $R^{(\pm)}_{Nq}(s x)= s^N R^{(\pm)}_{Nq}(s x)$ and $q$
enumerates different polynomials of the same degree. Several lowest  polynomials with the corresponding
anomalous dimensions can be found in Appendix~\ref{Appendix:Polynomials}. Here, we note that there exists only one
polynomial of the degree $N=0$,
$R_{00}^{(-)}=1$. Therefore, the expansion of the positive signature DAs, i.e $\Phi_4^{(B,A)}$
with $\varkappa(B,A)=+$, goes over the polynomials of degree $N\geq 1$.

 The genuine twist-$5$  distributions have a similar expansion:
\begin{align}
\Phi_{5}^{(B,A)g}(x) & = x_3\sum_{N\geq1,q} \zeta_{Nq}^{(B,A)}(\mu) \, T^{\varkappa(B,A)}_{Nq}(x)\,.
\end{align}
We emphasize, however, that neither the polynomials $T_{Nq}^{\pm}$ nor the anomalous dimensions have been calculated
so far.

The structure of the WW contributions to the DAs~\eqref{Phi4} and \eqref{Phi5} do not depend on the flavor content of
baryons and can be easily extracted from the results of Ref.~\cite{Anikin:2013yoa}.
We derive
\begin{align} \label{WW43}
\Phi_{4}^{(B,A)WW}(x) &=-x_1 x_2 \sum_{Nq}\frac{
\phi_{Nq}^{(B,A)}(\mu)}{(N+2)(N+3)}
\notag\\
&\quad \times
\Big[N+2-\partial_{x_3}\Big]  x_3\, P_{Nq}^{\varkappa(B,A)}(x)\,,
\end{align}
where %
summation goes over the polynomials of the parity $\varkappa(B,A)$.

For the twist-$5$ DAs, the WW terms of twist-$3$ and twist-$4$ have the  form
\begin{multline}
\Phi_5^{(B,A)WW_3}(x) =  x_3\sum_{Nq}\frac{
\phi^{(B,A)}_{Nq}(\mu)}{(N+2)(N+3)} \biggl[
-(N+2)^2
\\
+
(N+1-\partial_{x_2})(N+2-\partial_{x_1})%
\biggr]x_1x_2\, P_{Nq}^{\varkappa(B,A)}(x)\,
\end{multline}
and
\begin{align}
\Phi_5^{(B,A) WW_4}(x)& = x_3 \sum_{Nq}\frac{\eta_{Nq}^{(B,A)}(\mu)}{(N+1)(N+3)}
\notag\\
&\quad\times\left[N+1-\partial_{x_2}\right]\,
x_2 \,  R_{Nq}^{\varkappa(B,A)}(x_3,x_2,x_1)\,,
\end{align}
respectively.

\subsection{Chiral distribution amplitudes}\label{sect:chiral-tw45}
We now define the {\it chiral} three-quark operators as
\begin{align}
 \mathds{ O}^\downarrow_{4,\vec{m}}(\mathbf{z}) &=
q^\downarrow_{-,m_1}(z_1) q^\downarrow_{+,m_2}(z_2)  q^\downarrow_{+,m_3}(z_3)\,,
\notag\\
 \mathds{ O}^\uparrow_{5,\vec{m}}(\mathbf{z}) &=
q^\downarrow_{+,m_1}(z_1) q^\downarrow_{-,m_2}(z_2)  q^\downarrow_{-,m_3}(z_3).
\end{align}
They are constructed from the left-handed chiral quarks and they have the helicity $-1/2$ and $+1/2$, respectively. The DAs
related to these operators are defined by
\begin{align}
\vev{0|\mathds{O}^\downarrow_{4,\vec{m}}(\mathbf{z})|B}
 & =\frac{m_B}4(\mu\lambda)  (u_B^{\downarrow})_+
\int \mathcal{D}x \,e^{-ip_+\sum z_k x_k}\,\Xi_{4,\vec{m}}^{(B)}(x)\,,
\notag\\
\vev{0|\mathds{O}^\downarrow_{5,\vec{m}}(\mathbf{z})|B}
 & =-\frac{m_B}4(\mu\lambda)  (u_B^{\downarrow})_-
\int \mathcal{D}x \,e^{-ip_+\sum z_k x_k}\,\Xi_{5,\vec{m}}^{(B)}(x)\,.
\end{align}
For each twist, there are $7$ independent DAs which can be chosen as follows
\begin{equation}\label{XiFun}
\begin{tabular}{c c c c}
\hline\\[-3mm]
\hspace{10pt} $    \Xi_{t,\,uud}^{(p)}$ \hspace{10pt}&
                  \hspace{10pt} $    \Xi_{t,\,uus}^{(\Sigma^+)}$ \hspace{10pt}&
                          \hspace{10pt} $   \Xi_{t,\,ssu}^{(\Xi^0)}$ \hspace{10pt}&
                          \hspace{10pt} $    \Xi_{t,\,uds}^{(\Lambda)}$ \hspace{10pt}
\\[2mm]
\hline\\[-3mm]

           \hspace{10pt}  \hspace{10pt}&
                  \hspace{10pt} $    \Xi_{t,\,suu}^{(\Sigma^+)}$ \hspace{10pt}&
                          \hspace{10pt} $   \Xi_{t,\,uss}^{(\Xi^0)}$ \hspace{10pt}&
                          \hspace{10pt} $    \Xi_{t,\,sdu}^{(\Lambda)}$ \hspace{10pt}
\\[2mm]
\hline
\end{tabular}
\end{equation}
Similarly to the previous case, we denote the DA of the baryon $B$ in the $k$-th row (see the table) by
$\Xi^{(B,k)}_t$. The proton DAs coincide with the conventional definitions of Ref.~\cite{Braun:2000kw}, i.e.
$\Xi^{(p,1)}_t(x)= \Xi_t(x)$. First two  DAs in the second row, $\Xi^{(\Sigma^+,2)}_t(x)$ and
$\Xi^{(\Xi^0,2)}_t(x)$, are obviously symmetric under interchange of two last arguments, while the
$\Xi^{(\Lambda,2)}_t(x)$ is antisymmetric under  $x_2\leftrightarrow x_3$. The nucleon DA $\Xi_t^{(p,2)}$ is not
independent
\begin{align}
\Xi_t^{(p,2)}(x)=-\bigl(1+P_{23}\bigr)\, \Xi_t^{(p,1)}(x)\
\end{align}
and, therefore, it is not listed in the table.

The conformal wave expansion for the nucleon DA $\Xi_4^{(p,1)}$ was worked out in Ref.~\cite{Braun:2008ia}. The
eigenfunctions of the evolution kernel for the higher twist chiral three-quark operators are characterized by  parity (which is  $\varepsilon=\{1,e^{\pm i 2\pi/3}\}$)
under the combined cyclic permutations in the position and flavor spaces.
The conformal expansion of the nucleon chiral DA, $\Xi_4^{(p,1)}$, goes over the certain combination of
the eigenfunctions
with parity $e^{\pm i 2\pi/3}$.
The expansion
of the chiral  DAs for the other baryons involves the eigenfunctions of the parity
$\varepsilon=1$ as well. A more detailed discussion can be found in Ref.~\cite{Braun:2008ia}. Here, we present the final
results only. The expansions for
$\Lambda$ baryon as compared to  all other baryons in the octet have a slightly different form. Namely,
\begin{align}
\Xi_4^{(B\neq\Lambda,k)}(x)&=x_2x_3 \left[\sum_{Nq} \xi^{(B)}_{Nq}(\mu) \Pi^{(k)}_{Nq}(x)
+\sum_{Nq} \theta_{Nq}^{(B)}(\mu)  { \Pi}^{(+)}_{Nq}(x)\right],
\notag\\
\Xi_4^{(\Lambda,k)}(x) & =x_2x_3 \left[\sum_{Nq} \xi^{(\Lambda)}_{Nq}(\mu)\Lambda^{(k)}_{Nq}(x)
\pm\sum_{Nq} \theta_{Nq}^{(\Lambda)}(\mu)  \Lambda^{(-)}_{Nq}(x)\right].
\end{align}
In the last formula, the signs $``\pm"$ correspond to $k=1,2$, respectively.
The polynomials $\Pi^{(k)}_{Nq}(x)$,
$\Lambda^{(k)}_{Nq}(x)$ are related to the eigenfunctions
$\Pi_{Nq}(x)$ of the parity $\epsilon=e^{\pm i 2\pi/3}$, which were calculated in Ref.~\cite{Braun:2008ia},
\begin{align}
\Pi^{(1)}_{Nq}(x) & =\Pi_{Nq}(x),
\notag\\
\Pi^{(2)}_{Nq}(x) & =-(1+P_{23})\Pi_{Nq}(x),
\notag\\
\Lambda^{(1)}_{Nq}(x) & =(1+2P_{23})\Pi_{Nq}(x),
\notag\\
\Lambda^{(2)}_{Nq}(x) & =-(1-P_{23})\Pi_{Nq}(x),
\end{align}
where $P_{23}$ is the permutation operator:
\begin{align}
P_{23} f(x_1,x_2,x_3)=f(x_1,x_3,x_2).
\end{align}
 The eigenfunctions $\widetilde \Pi^{(+)}_{Nq}$
and $ \Lambda^{(-)}_{Nq}$ belong to the parity sector $\epsilon=1$ and they are (anti)symmetric under
$x_2\leftrightarrow x_3$  permutation:
\begin{align}
P_{23}\, \Pi^{(+)}_{Nq}(x)=\Pi^{(+)}_{Nq}(x)\,, &&P_{23}\, \Lambda^{(-)}_{Nq}(x)=-\Lambda^{(-)}_{Nq}(x)\,.
\end{align}
Notice that for the proton $\theta_{Nq}^{(p)}(\mu)=0$ by symmetry.

In order to write down the WW contributions to the twist-$5$ amplitudes, we define the following functions
\begin{align}
\label{fun1}
W^{(B)}(x) & =- P_{12}\sum_{Nq} \frac{\xi^{(B)}_{Nq}(\mu)}{(N+1)(N+3)}
\biggl[(N+1-\partial_3) x_3 \, \Pi^{(1)}_{Nq}(x)
\notag\\
&\quad
+(N+1-\partial_1) x_1  \Pi^{(2)}_{Nq}(x)\Biggr],
\notag\\
W^{(\Lambda)}(x) & = P_{12}\sum_{Nq} \frac{\xi^{(\Lambda)}_{Nq}(\mu)}{(N+1)(N+3)}
\biggl[(N+1-\partial_3) x_3 \, \Lambda^{(1)}_{Nq}(x)
\notag\\
&\quad
+(N+1-\partial_1) x_1  \Lambda^{(2)}_{Nq}(x)\Biggr],
\end{align}
and
\begin{align}
\label{fun2}
W^{(B)}_+(x) & =- P_{12}\sum_{Nq} \frac{\theta^{(B)}_{Nq}(\mu)}{(N+1)(N+3)}
\biggl[(N+1-\partial_3) x_3 \,
\notag\\
&\quad
+(N+1-\partial_1) x_1 \Biggr] \Pi^{(+)}_{Nq}(x),
\notag\\
W^{(\Lambda)}_-(x) & = P_{12}\sum_{Nq} \frac{\xi^{(\Lambda)}_{Nq}(\mu)}{(N+1)(N+3)}
\biggl[(N+1-\partial_3) x_3 \,
\notag\\
&\quad
+(N+1-\partial_1) x_1  \Biggr]  \Lambda^{(-)}_{Nq}(x),
\end{align}
With the help of Eqs.~(\ref{fun1}) and (\ref{fun2}), the WW contributions to the twist-$5$ DA can be written in the
following form
\begin{align}
\Xi_5^{(B,1)WW}(x) & =x_1\Big(W^{(B)}(x)+W^{(B)}_+(x)\Big)\,,
\notag\\
\Xi_5^{(B,2)WW}(x) & =x_1\Big(-(1+P_{23})W^{(B)}(x)+W^{(B)}_+(x)\Big)\,,
\notag\\
\Xi_5^{(\Lambda,1)WW}(x) & = x_1\Big(W^{(\Lambda)}(x)+W^{(\Lambda)}_-(x)\Big)\,,
\notag\\
\Xi_5^{(\Lambda,2)WW}(x) & = x_1\Big((1-P_{23})W^{(\Lambda)}(x)- W^{(\Lambda)}_-(x)\Big)\,.
\end{align}
The first few polynomials $\Pi_{Nq}$, $\Pi^{(+)}_{Nq}$, $\Lambda_{Ng}^{(-)}$ are given in
Appendix~\ref{Appendix:Polynomials}.

\section*{Acknowledgements}
We are grateful to V.~M.~Braun for useful discussions.
The work was supported by the DFG (A.M), grant
MO~1801/1-1, and by DAAD (I.A).


\appendix
\renewcommand{\theequation}{\Alph{section}.\arabic{equation}}

\section*{Appendices}

\section{Isospin relations}\label{Appendix:match}
The isospin symmetry allows one to express DAs for any baryon in the octet in terms of the DAs  discussed in the
main part of the paper. We remind that the DAs
           $\Phi_{t,\vec{m}}^{(B)}$, $\Xi_{t,\vec{m}}^{(B)}$
depend on the flavor indices
            $m_1, m_2, m_3$
of the quark fields in the corresponding light-ray operators, see Eqs.~\eqref{defPhi3}, \eqref{defPhi4} and
\eqref{Phi5}.

The independent DAs for $p,\, \Sigma^+,\Xi^0,\Lambda$ baryons are collected in the Tables~\eqref{Our-Fun},
\eqref{Fun-45} and \eqref{XiFun}. Defining distribution amplitudes  for the remaining members of the octet we have
to specify the  flavor indices of the corresponding DAs. For the $\Sigma^0$ baryon DAs we simply take over the
flavor indices of $\Lambda$ baryon. The $n,\Sigma^-, \Xi^-$ baryon DAs  inherit flavor indices from the
corresponding DAs for $p,\Sigma^+, \Xi^0$ baryons with the substitution $u\leftrightarrow d$.
For example, the  twist-$3$ DAs that have to be added to the first row of the table~\eqref{Our-Fun} are:
         $\Phi_{3,ddu}^{(n)}$, $\Phi_{3,dds}^{(\Sigma^-)}$,  $\Phi_{3,ssd}^{(\Xi^-)}$ and
         $\Phi_{3,uds}^{(\Sigma^0)}$.

We accept the same phase conventions for the baryons in octet as in Ref.~\cite{Bali:2015ykx} (see Appendix A there).
Then the following relations  hold
\begin{align}
&\text{DA}^{(p)}(x) =-\text{DA}^{(n)}(x)\,,\notag\\
&\text{DA}^{(\Xi^-)}(x) =-\text{DA}^{(\Xi^0)}(x)\,,\notag\\
&\text{DA}^{(\Sigma^+)}(x) =-\text{DA}^{(\Sigma^-)}(x)=-\sqrt{2}\text{DA}^{(\Sigma^0)}(x)\,,
\end{align}
between the DAs of the same type, e.g. (see table~\eqref{Fun-45})
\begin{align*}
\Phi^{(\Sigma^+)}_{5, uus}(x)=-\Phi^{(\Sigma^-)}_{5, dds}(x)=-\sqrt2\Phi^{(\Sigma^0)}_{5, uds}(x)\,.
\end{align*}
Below we present the relations between the functions used in the present paper and those introduced in Ref.~\cite{Bali:2015ykx}:
\begin{align}
\label{Com-DA}
\Phi_{\text{\scriptsize\cite{Bali:2015ykx}}
\pm}^{B\neq\Lambda}(x) &=\frac12(1+P_{13})\Phi_3^{(B, 1)}(x)\,,
\notag\\
\Pi_{\text{\scriptsize\cite{Bali:2015ykx}}}^{B\neq\Lambda}(x) & =-\frac12\Phi_3^{(B, 2)}(x)\,,
\end{align}
where $P_{13}$ is  the permutation operator and
\begin{align}
\label{Com-Lambda}
\Phi_{\text{\scriptsize\cite{Bali:2015ykx}}
+}^{\Lambda}(x) &=\sqrt{\frac16}(1+P_{13})\Phi_3^{(\Lambda, 1)}(x)\,,
\notag\\
\Phi_{\text{\scriptsize\cite{Bali:2015ykx}}
+}^{\Lambda}(x) &=-\sqrt{\frac32}(1-P_{13})\Phi_3^{(\Lambda, 1)}(x)\,,
\notag\\
\Pi_{\text{\scriptsize\cite{Bali:2015ykx}}}^{\Lambda}(x) & =-\sqrt{\frac32}\Phi_3^{(\Lambda, 2)}(x)\,.
\end{align}
The normalization factors in Eqs.~\eqref{Com-DA}, \eqref{Com-Lambda} ensure that in the limit of $SU(3)$ flavor
symmetry the following relations hold~\cite{Wein:2015oqa,Bali:2015ykx}
\begin{align}
&\Phi_{\text{\scriptsize\cite{Bali:2015ykx}}
+}^{p}
=\Phi_{\text{\scriptsize\cite{Bali:2015ykx}}+}^{\Sigma^-}
=\Phi_{\text{\scriptsize\cite{Bali:2015ykx}}+}^{\Xi^0}
=\Phi_{\text{\scriptsize\cite{Bali:2015ykx}}+}^{\Lambda}
=\Pi_{\text{\scriptsize\cite{Bali:2015ykx}}}^{\Sigma^-}
=\Pi_{\text{\scriptsize\cite{Bali:2015ykx}}}^{\Xi^0}
\notag\\
&\Phi_{\text{\scriptsize\cite{Bali:2015ykx}}-}^{p}
=\Phi_{\text{\scriptsize\cite{Bali:2015ykx}}-}^{\Sigma^-}
=\Phi_{\text{\scriptsize\cite{Bali:2015ykx}}-}^{\Xi^0}
=\Phi_{\text{\scriptsize\cite{Bali:2015ykx}}-}^{\Lambda}
=\Pi_{\text{\scriptsize\cite{Bali:2015ykx}}}^{\Lambda}\,.
\end{align}
%

\section{Polynomials}\label{Appendix:Polynomials}
In this  Appendix, we list lowest polynomials which enter the expansion of the baryon DAs of different twists. We also give the
anomalous dimensions of the coefficients that accompany these polynomials.

The twist-$3$ polynomials have the forms:
\begin{align}
P_{00}(x) &= 1\,,
\notag\\
P_{10}(x) &= (x_1-x_3)/2\,,
\notag\\
P_{11}(x) &= (x_1+x_3-2 x_2)/2\,,
\notag\\
P_{20}(x) &= 3x^2_1 - 3x_1x_2 + 2x^2_2 - 6x_1x_3 - 3x_2x_3 + 3x^2_3\,,
 \notag\\
P_{21}(x) &= (x_1-x_3)(x_1+x_3-3x_2)\,,
\notag\\
P_{22}(x) &=x_1^2+x_3^2-12 x_1 x_3+ 9 x_2 (x_1+x_3)-6x_2^2.
\end{align}
The corresponding one loop  anomalous dimensions are (in units of $\alpha_s/2\pi$)
\begin{align}
\gamma_{Nq}=\Big\{\frac23, \ \frac{26}9,\  \frac{10}3,\  \frac{38}9,\  \frac{46}9,\  \frac{16}3\Big\}.
\end{align}
The mixed chirality twist-$4$ polynomials are given by
\begin{align}
R_{00}^-(x) &= 1\,,
\notag\\
R_{10}^-(x) & = x_1+x_3-\frac32 x_2\,,
\notag\\
R_{10}^+(x) & = x_1-x_3-\frac12 x_2\,,
\notag\\
R_{20}^+(x) & = x_2^2-x_3^2 - 2 x_1 x_2+3x_1x_3-\frac23 x_1^2\,,
\notag\\
\begin{pmatrix}
R_{20}^-(x)\\
R_{21}^-(x)
\end{pmatrix}  &= x_2^2+\frac49\left(-5\pm\sqrt{43}\right)x_2 x_3+x_3^2 \notag\\
  &\quad +\frac29\left(1\mp2\sqrt{43}\right)x_1x_2-\frac19\left(17\pm2\sqrt{43}\right)x_1x_3
  \notag\\
  &\quad +\frac4{27}\left(4\pm\sqrt{43}\right)x_1^2\,.
\end{align}
The corresponding anomalous dimensions are~\cite{Braun:2008ia}
\begin{align}
\gamma_{Nq} &=\Big\{-2, \ \frac{2}9,\ 2,\  \frac{32}9,\  \frac{2(14-\sqrt{43})}9,\  \frac{2(14+\sqrt{43})}9\Big\}.
\end{align}

The chiral  twist-$4$ polynomials take the form
\begin{align}
\Pi_{00}(x) &=1\,,\notag\\
\Pi_{10}(x) &=x_1+x_3-\frac32 x_2\,,\notag\\
\Pi_{20}(x) &=x_1^2-4x_1 x_2 +2 x_2^2+2x_1 x_3-4x_2 x_3+x_3^2\,.
\end{align}
These polynomials correspond to $\gamma_{Nq}=\{-2, 4/3, 4/3 \}$. Finally, for the polynomials $\Lambda^\pm$ and $\Pi^+$ we get
\begin{align}
\Lambda_{10}^{(-)}(x)& =x_2-x_3\,,
\notag\\
\Lambda_{20}^{(-)}(x) &=x_2^3-x_3^2-6x_1(x_2-x_3)\,,
\\
\Pi_{20}^{(+)}(x) &=x_1^2+\frac1{3}(x_2^3+x_3^2)-\frac12 x_2 x_3- x_1(x_2+x_3)\,
\notag
\end{align}
with the anomalous dimensions $\gamma_{Nq}=\{-2/3, 4, 4\}$.
\vskip 5mm

We also write down here  the first nonzero terms of the WW contributions  for the  mixed chirality DAs. For the
twist-$4$ DAs one gets
\begin{align*}
\Phi_{4}^{(B,A)WW}(x) &= \frac16\,\phi_{00}^{(B,A)}\,x_1x_2 (1-2x_3)+\ldots\\
\Phi_{4}^{(B,A)WW}(x) &= \frac{\phi_{10}^{(B,A)}}{24}
x_1x_2
\Big(x_1(1-3x_3)-x_3(2-3x_3)\Big)+\ldots
\end{align*}
for the positive and negative signature DAs, respectively.

For the twist-$5$ functions one gets
\begin{align*}
\Phi_{5}^{(B,A)WW_3}(x) & = \frac{\phi_{00}^{(B,A)}}6 x_3\big (1-2x_1-x_2(1+2x_1)\big)+\ldots
\notag\\
\Phi_{5}^{(B,A)WW_4}(x) & = \frac{\eta_{10}^{(B,A)}}3 x_3\big (1-2x_3-x_2(1+x_1-3x_3)\big)+\ldots
\end{align*}
for the positive signature and
\begin{align*}
\Phi_{5}^{(B,A)WW_3}(x) &=\frac1{12}\phi_{10}^{(B,A)} x_3\Big(x_1-\frac12 x_3+x_2(x_3-2x_1)\notag\\
&\quad +\frac32x_1(x_3-x_1)+\frac32x_1x_2(x_3-x_1)\Big)+\ldots
\\
\Phi_{5}^{(B,A)WW_4}(x) &= -\frac13\,\eta_{00}^{(B,A)}\,x_3\big (1-x_2\big)+\ldots
\end{align*}
for the negative signature DAs.

\end{document}